\documentstyle[aps,epsfig,manuscript]{revtex}

\newcommand{\Nav}{\mbox{$\langle N \rangle$}}

\newcommand{\bb}{\mbox{$\left< l^2 \right>$}}

\newcommand{\Rend}{\mbox{$R_e$}}
\newcommand{\Rgyr}{\mbox{$R_g$}}

\newcommand{\nuD}{\mbox{$\nu_{I}$}}
\newcommand{\nuS}{\mbox{$\nu_{II}$}}

\newcommand{\alphaD}{\mbox{$\alpha_{I}$}}
\newcommand{\alphaS}{\mbox{$\alpha_{II}$}}
\newcommand{\alphaM}{\mbox{$\alpha_{eff}$}}
\newcommand{\deltaD}{\mbox{$\delta_{I}$}}
\newcommand{\deltaS}{\mbox{$\delta_{II}$}}

\newcommand{\phistar}{\mbox{$\phi^{*}$}}
\newcommand{\Nstar}{\mbox{$N^{*}$}}

\newcommand{\fchain}{\mbox{$f_{chain}$}}
\newcommand{\fend}{\mbox{$f_{end}$}}

\newcommand{\gL}{\mbox{$f_1$}}
\newcommand{\cR}{\mbox{$c_0$}}
\newcommand{\cL}{\mbox{$c_1$}}

\begin{document}
\title{Dynamical Monte Carlo Study of Equilibrium Polymers (II): 
Effects of High Density and Ring Formation}
\author{A. Milchev$^1$, J.P. Wittmer$^2$\thanks{%
email:jwittmer@dpm.univ-lyon1.fr}, and D.P. Landau$^3$}
\address{$^{1}$ Institute for Physical Chemistry, Bulgarian Academy of Sciences,\\
$1113$ Sofia, Bulgaria\\
$^2$ D\'epartment de Physique des Mat\'eriaux, Universit\'e Lyon I \& CNRS,\\
69622 Villeurbanne Cedex, France.\\
$^3$ Department of Physics and Astronomy, University of Georgia,\\
Athens, Ga. 30602, U.S.A.}
\maketitle

\begin{abstract}
An off-lattice Monte Carlo algorithm for solutions of equilibrium polymers
(EP) is proposed. At low and moderate densities this is shown
to reproduce faithfully the (static) properties found recently for
flexible linear EP using a lattice model \cite{WMC98}. The molecular weight
distribution (MWD) is well described in the dilute limit by a Schultz-Zimm
distribution and becomes purely exponential in the semi-dilute limit.

Additionally, very concentrated molten systems are studied. 
The MWD remains a pure exponential in contrast to recent claims\cite{Yannick}.
The mean chain mass is found to increase faster with density than in the
semi-dilute regime due to additional entropic interactions generated by the
dense packing of spheres.

We also consider systems in which the formation of rings is allowed so that
both the linear chains and the rings compete for the monomers. In agreement
with earlier predictions the MWD of the rings reveals a strong singularity
whereas the MWD of the coexisting linear chains remains essentially
unaffected.
\end{abstract}

\sloppy         

\centerline{\today}

\vskip 1.0truecm \centerline{PACS numbers: 82.35.+t, 61.25H, 64.60C }

\section{Introduction}


The molecular mass distribution (MWD) of systems of linear unbranched
``equilibrium polymers'' (EP) is essentially exponential \cite{WMC98}. 
In EP systems polymerization takes place under condition of chemical 
equilibrium between polymer chains and their respective monomers. 
A classical example we will focus on is provided
by systems of surfactants forming polydisperse solutions of long worm-like
aggregates, called ``giant micelles'' (GM), which combine with each other,
or break into smaller fractions \cite{com:GM}.


Despite polydispersity, EP resemble in many aspects conventional quenched
polymers where the polymerization reaction has been deliberately terminated.
Recently, the basic scaling predictions for EP \cite{com:GM} based on
classical polymer physics \cite{Degennesbook} have been tested by two of us
(AM, JPW) \cite{WMC98} by means of a lattice Monte Carlo simulation based on
the widely used Bond-Fluctuation Model (BFM) \cite{BFM}. This demonstrated
excellent agreement with theory over a very broad range of density and
temperature variation. Specifically, it was shown that the MWD takes the
form, 
\begin{equation}
c(x)dx\propto \left\{ 
\begin{array}{ll}
\exp (-x)dx & \mbox{($\Nstar \gg \Nav, \phistar \ll \phi$)} \\ 
x^{\gamma -1}\exp (-\gamma x)dx & 
\mbox{($\Nstar \ll \Nav, \phistar \gg
\phi$)}.
\end{array}
\right.  \label{eq:MWD_GM}
\end{equation}
The scaling parameter $x=N/\Nav$ is the ratio of the
chain mass $N$ and the mean mass \Nav, 
\Nstar\ and \phistar\ mark the mean mass and the density at the crossover 
from dilute to semi-dilute regimes at given scission energy $E$, and 
$\gamma \approx 1.158$ is the susceptibility exponent of the $n\rightarrow 0$
vector model in 3D \cite{com:gamma}. 
The mean chain length \Nav\ was confirmed to vary with density $\phi $ 
and the (dimensionless) scission energy $E$ as 
\begin{equation}
\Nav\approx \Nstar \left( \phi /\phistar\right)^{\alpha }\propto \phi^{\alpha}\exp(\delta E)
\label{eq:Nav}
\end{equation}
with exponents $\alphaD=\deltaD=1/(1+\gamma )\approx 0.46$ in the dilute and 
$\alphaS=(1+(\gamma -1)/(3\nuD -1))/2\approx 0.6$, 
$\deltaS=1/2$ in the semi-dilute regime.
The exponent $\nuD \approx 0.588$ is the swollen chain (self-avoiding walk)
exponent in 3D.

Recently, these results have been questioned in an interesting computational
study where a reptation algorithm was used on a cubic lattice \cite{Yannick}.
It was suggested that the MWD becomes singular $c(N)\propto N^{-\tau }$
with $\tau \approx 0.56$ at very high volume fractions of order unity. This
finding, if corroborated, might be of some relevance in view of the alleged
Levy-flight dynamic behavior in system of GM \cite{Levyflight}. This claim
prompted the present off-lattice Monte Carlo (OLMC) approach to be applied
to the high density limit which was not accessible within our previous
approach. We stress that effects only occurring at volume fractions of order
one are unlikely to be relevant for real GM, but would be interesting on
more general theoretical grounds in that the MWD in fact probes the free
energy of the dense packing of beads.

As we are going to show, new physics (i.e. additional terms in the system
free energy) intervenes indeed due to packing effects of the spherical
beads. This increases effectively the growth exponent $\alpha $ for \Nav,
but does not effect the scaling form of the MWD. 
We do not observe any trace of singularity --- 
in perfect agreement with the analytical predictions \cite{com:GM,SC94}.

The investigation of the high density limit is only one motivation for the
off-lattice algorithm proposed. In addition to this an effective OLMC for EP
algorithm is highly warranted to overcome the usual shortcomings of lattice
models and to serve in examining the role of polymers semi-flexibility. It
is also a better tool in dynamic studies of a broader class of soft
condensed matter systems where bifunctionality of the chemical bonds might
be extended to polyfunctional bonds, as this is the case in gels and
membranes. Note that the OLMC was already applied successfully to
rheological properties of EP reported in \cite{MWL99a} and to systems of
EP brushes \cite{MWL99b}.

The scope of this paper is three-fold: we want to present the OLMC scheme
and to test it by comparing the new results with previous findings (in the
dilute and semi-dilute regime) obtained by means of our lattice Monte Carlo
approach described in \cite{WMC98}. Secondly, we wish to address the physics
in the melt density regime. In addition, the OLMC algorithm is tested in
systems in which the formation of closed loops is allowed. We demonstrate
for the first time in a computer experiment that there is a singularity of
the ring MWD $\cR(N)\propto N^{-\tau }$ where $\tau =5/2$ in 3D \cite{Porte,SW99}.

%
%
This paper is organized as follows. 
After presenting the algorithm in Sec.\ref{sec:Algo}
we focus in Sec.\ref{sec:RN} on the different density and chain length
regimes reflected, e.g., by the distribution of the radius of gyration 
$\langle R_{g}^{2}(N)\rangle$ versus mass $N$ and other conformational
properties. We discuss subsequently the MWD in Sec.\ref{sec:MWD}. There we
compare our OLMC results for mixed systems (in which rings are also present)
with data obtained with the BFM and with a Grand Canonical lattice algorithm
based on the mapping of the EP problem on a Potts model. The scaling of the
mean chain mass is considered in Sec.\ref{sec:Nav}. 
The theoretical concepts (which have already been extensively considered elsewhere 
\cite{com:GM,SC94}) are briefly discussed {\em en passant}. 
We show that agreement of the OLMC with our previous work is excellent for
small and moderate densities. 
For volume fractions larger $0.1$ we evidence a third molten regime. 
In the final Section\ref{sec:Conclusion} we summarize our findings.

\section{The Computational Algorithm}\label{sec:Algo}

\begin{figure}[tbp]
\centerline{
\epsfig{file=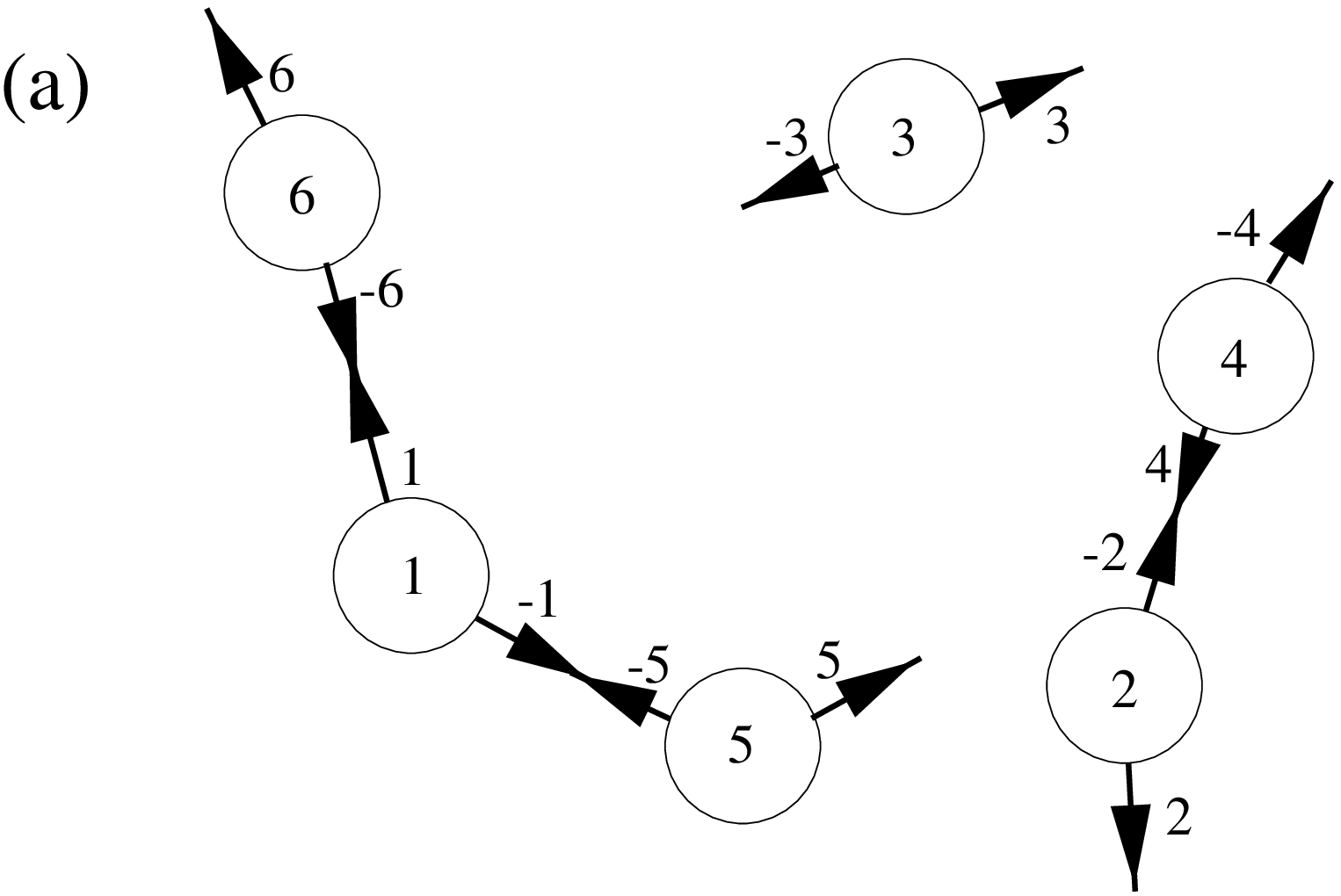,width=90mm,height=70mm}
\epsfig{file=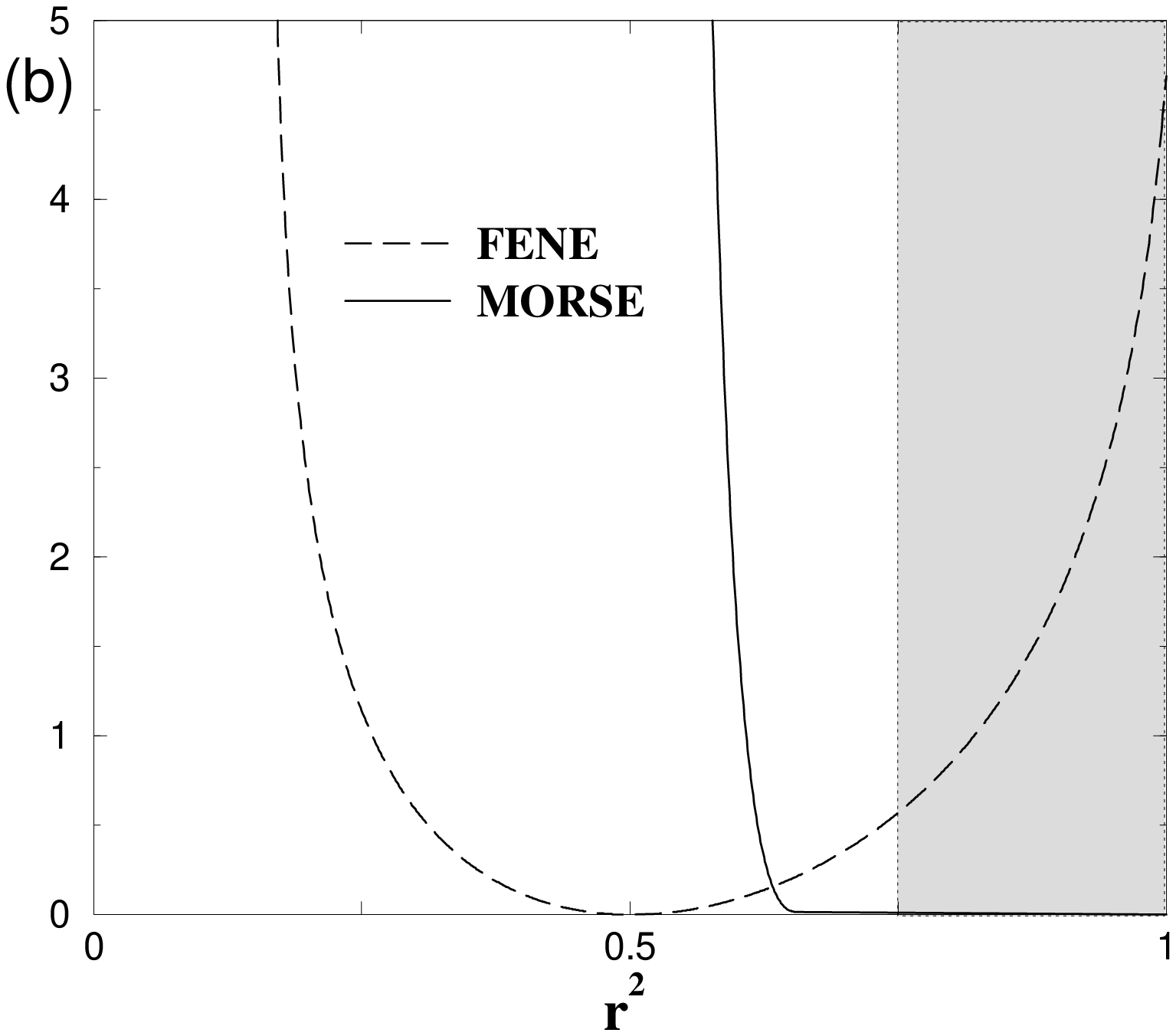,width=80mm,height=70mm}}
\vspace*{0.5cm}
\caption{Sketch of algorithm: (a) Bonds between
spherical beads break and recombine constantly with rates depending on the scission 
energy $E$ which is assumed independent of mass $N$ and number density $\phi$.
Each monomer has two (saturated or unsaturated) bonds. 
Chains consists of symmetrically connected lists of bonds. 
The data structure is based on the bonds rather than the polymer chains. 
(b) Plots of bonded (FENE) and non-bonded (Morse) interactions used
in the present model. The shaded area denotes distances where scission -
recombination events may take place. }
\label{fig:algo}
\end{figure}

It is clear that in a system of EP where scission and recombination of bonds
constantly take place the particular scheme of bookkeeping should be no
trivial matter \cite{com:Potts}. Since chains are only transient objects
the data structure of the chains can only be based
on the individual monomers, or, even better, on the saturated or unsaturated
bonds \cite{WMC98}. This idea is depicted in Fig.\ref{fig:algo}(a). Each
bond is considered as a {\em pointer}, originating at a given monomer and
pointing to the respective other bond with which the couple forms a nearest
neighbor, or to $NIL$ (nowhere), if the bond is free (unsaturated). Chains
consists of symmetrically connected lists of bonds: $jbond=pointer(ibond)
\leftrightarrow ibond=pointer(jbond)$. Recombination of the two initially
unsaturated bonds $ibond=2$ and $jbond=5$ connects the respective monomers 
$imon=2$ and $jmon=5$ in Fig.\ref{fig:algo}(a). Note that only two pointers
have to be changed and that the remaining chains behind both monomers are
not involved. Breaking a saturated bond $ibond$ requires resetting the
pointers of the two connected bonds $ibond$ and $jbond=pointer(ibond)$ to $NIL$.

As mentioned in the Introduction this data structure has been incorporated 
\cite{WMC98} within the widely used BFM algorithm \cite{BFM}. For the
off-lattice version presented here we have now harnessed a very efficient
bead-spring algorithm for polymer chains (for technical details see \cite
{OLMCold}) and cast it onto the data structure described above. This
off-lattice Monte Carlo (OLMC) scheme is characterized by the bonded and the
non-bonded interactions shown in Fig.\ref{fig:algo}(b).


Each bond is described by a shifted {\sc FENE} potential where a bond of
length $r$ has a maximum at $r_{max}=1$ 
\begin{eqnarray}
U_{FENE}(r)=-K(r_{max}-r_0)^2\ln \left [ 1- \left (\frac{r-r_0}{r_{max}-r_0}%
\right )^2 \right ] - E.  \label{eq:FENE}
\end{eqnarray}
$E$ corresponds to a constant scission energy.
Note that \mbox{$U_{FENE}(r=r_0)=-E$} and that $U_{FENE}$
near its minimum at $r_0$ is harmonic, with $K$ being the spring constant,
and the potential diverges logarithmically to infinity both when 
$r\rightarrow r_{max}$ and $r\rightarrow r_{min}=2r_0-r_{max}$. 
Thus the FENE
potential does not need to be truncated at $r_{min}$ and $r_{max}$ and
discontinuities in the derivative of the potential are avoided. Following
ref.\cite{OLMCold} we choose the parameters $r_{max}-r_0=r_0-r_{min}=0.3$
and $K/T=40$, $T$ being the absolute temperature. The units are such that
the Boltzmann's constant $k_B = 1$.


The non-bonded interaction between effective monomers is described by a
Morse-type potential, $r$ being the distance between the beads 
\begin{eqnarray}
U_M(r)= \exp[-2a(r-r_{min})]-2\exp[-a(r-r_{min})]  \label{eq:Morse}
\end{eqnarray}
with parameters $a=24$ and $r_{min}=0.8$. The latter parameter sets roughly
the sphere diameter. The $\Theta$-temperature of the coil-globule transition
for our model is $\Theta \approx 0.62$ so that at $T=1$ we work under good
solvent conditions \cite{OLMCold}.


The model can be simulated fairly efficiently with a dynamic MC algorithm,
as described previously\cite{OLMCold}. The trial update involves choosing a
monomeric unit at random and attempting to displace it randomly by
displacements $\Delta x, \Delta y, \Delta z$ chosen uniformly from the
interval $-0.5\leq \Delta x, \Delta y, \Delta z\leq 0.5$. Moves are then
accepted according to the Metropolis criterion and one Monte Carlo step
(MCS) involves as many attempted moves as there are monomers in the system. 


In equilibrium polymers the bonds between neighbors along the backbone of a
chain are constantly subject to scission and recombination events. In the
present model only bonds, stretched a distance $r$ beyond some threshold
value, $r_b$, are attempted to break so that eventually an energy $%
U_{FENE}(r) > 0$ could be released if the bond breaks. Since each monomer
may have at most two bonds in the same time, all particles with unsaturated
bonds (two for single monomers and one for chain ends) may form new bonds,
once they approach each other within the same interval of distances $r_b \le
r \le 1$ where scissions take place. (Note that recombination for $r < r_b$
would violate detailed balance.)

%

We do not allow in the presented study for branching. However, more than two
bonds per monomer are possible in principle and this feature may readily
included in the algorithmic framework. The generalization on netted
structures like membranes or sponges is evident. (Note that this is less
straightforward in the BFM scheme due to its lattice character which
generates ergodicity problems.) Obviously, another big advantage of the
off-lattice scheme compared to its lattice precursor is its applicability to
rheological problems \cite{MWL99b}. 


In most parts of this paper we focus on systems where no formation of ring
polymers is allowed. This condition has to be observed whenever an act of
polymerization takes place. Because there is no direct chain information in
the data structure this has to be done by working up the list of pointers
(which adds only four lines to the source code). In physical time units the
simulation becomes {\em faster} for higher $E$: the number of recombinations
per unit time goes like $\exp (-E)$, but the chain mass only grows as $\exp
(E/2)$. Obviously, the algorithm becomes even faster for the mixed systems
discussed in Sec.\ref{subsec:rings} where the ring closure constraint has
been dropped.

\section{Conformational properties}\label{sec:RN}

The presented algorithm allows us to simulate a large number of 
particles at very modest expenses of operational memory. 
Most of the results in the present study involve $65536$ particles 
for number densities between $\phi=0.125$ up to $\phi =1.5$.
%
Note that our highest densities correspond to very concentrated solutions.
This can be better seen from the volume fractions which vary between 
$v\phi\approx 0.03$ and $0.33$.
The latter value has to be compared with the (only slightly larger) hard-sphere 
freezing volume fraction (``Alder transition") of about one half \cite{Pusey}
--- it correponds, in fact, to a relatively dense hard-sphere liquid.
Note also that dense globules of maximal $\phi\approx 2$ are formed by quenched 
monodisperse polymers in bad solvent \cite{OLMCvirial}. 

The effective bead volume $v\approx \pi l^{3}/6\approx 0.22$ used above was 
estimated with the measured mean bond length $l\approx 0.75$ from Tab.\ref{tab:phiE}. 
(Note that $l$ is weakly decreasing with density \cite{OLMCold} and 
that it is slightly smaller than the ``bead diameter" given by the shifted 
Morse parameter $r_{min}$.) A similar, marginally larger volume of $v\approx 0.25$ 
may be obtained from the virial expansion for quenched polymers \cite{OLMCvirial}.
%

%
%
The scission energy $E$ was varied in a similar range as in \cite{WMC98}
from $E=4$ up to $E=12$ to yield sufficiently strong chain mass variation.
Note that the \Nav\ remains always two
orders of magnitudes smaller than the total particle number within the box.
Hence, from our previous study \cite{WMC98} one expects finite box-size
effects to be small. This was indeed born out by finite-size test performed
by varying the box sizes from $16^{3}$ over $32^{3}$ to $64^{3}$. Only the
results from the largest boxes simulated for a configuration $(E,\phi )$ are
reported here.


Periodically the whole system is examined, the MWD $c(N)$ 
(discussed in the subsequent section) and the distributions
of the squared end-to-end distance $\langle R_{e}^{2}(N)\rangle$ and
gyration radius, $\langle R_{g}^{2}(N)\rangle$ (averaged over all chains
of a {\em particular} mass $N$) are counted and stored. 
(Periodic boundary conditions are implemented and interactions between 
monomers follow the minimum image convention. The computation of the
conformational properties of the chains then imply a restoration of 
{\em absolute} monomer coordinates from the periodic ones for each 
repeating unit of the chain.)
Moments obtained from these distributions are presented in 
Tab.\ref{tab:phiE} and Tab.\ref{tab:phi}.

\begin{figure}[tbp]
\centerline{\epsfig{file=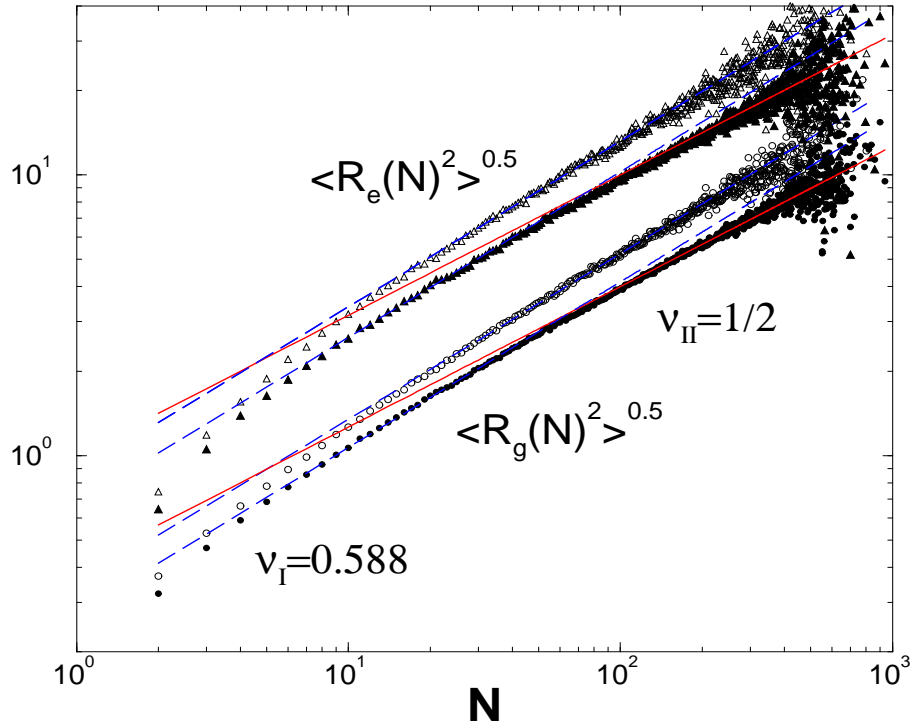,width=120mm,height=100mm}}
\caption{Variation of the end-to-end distance \Rend\ (triangles) and the 
radius of gyration \Rgyr\ (squares) with chain mass $N $ at $E=7$ for dilute 
(open symbols: $E=7,\phi =0.125$) and
concentrated (full symbols: $E=7,\phi=1$) systems. The dashed
lines denote the dilute exponent $\nuD=0.588$ which is visible
for the dilute systems over the full range of $N$ and for the semi-dilute
system for small $N$, i.e. within the blob. The slope $\nuS=1/2$
indicates the Gaussian statistics for strongly overlapping long chains.}
\label{fig:RN}
\end{figure}

The different simulational regimes are most easily demonstrated by the
conformational changes with density and chain mass. We demonstrate that
the conformational properties for EP follow the same universal functions as
conventional quenched polymers.
In Fig.\ref{fig:RN} $\langle R_{e}^{2}(N)\rangle^{1/2}$ and 
$\langle R_{g}^{2}(N)\rangle^{1/2}$ are plotted versus $N$.
One configuration in the dilute regime and one in the concentrated
limit (both at same $E=7$) have been presented. In the first case one
clearly sees the swollen coil exponent. 
From the upper dashed line the persistence length of swollen 
EP is estimated as $b=\Rend/N^{0.588}\approx 0.92$.
In the latter case only chains smaller than the excluded volume blob of size 
$\xi$ size are swollen. Larger chains ($N\gg g$) show Gaussian behavior with 
$\Rend/\sqrt{6}\approx \Rgyr \propto N^{1/2}$.
This slope was used to estimate the variation of the effective persistence length 
$b=\Rend/N^{0.5}$ with $\phi$. As can be seen from Tab.\ref{tab:phi} $b$ decreases
with increasing density, hence with decreasing blob size $\xi$ which we have
estimated directly from the intercept of the two slopes at the radius of gyration. 
Note that $\xi$ is relatively small for the densities computed.  

\begin{figure}[tbp]
\centerline{\epsfig{file=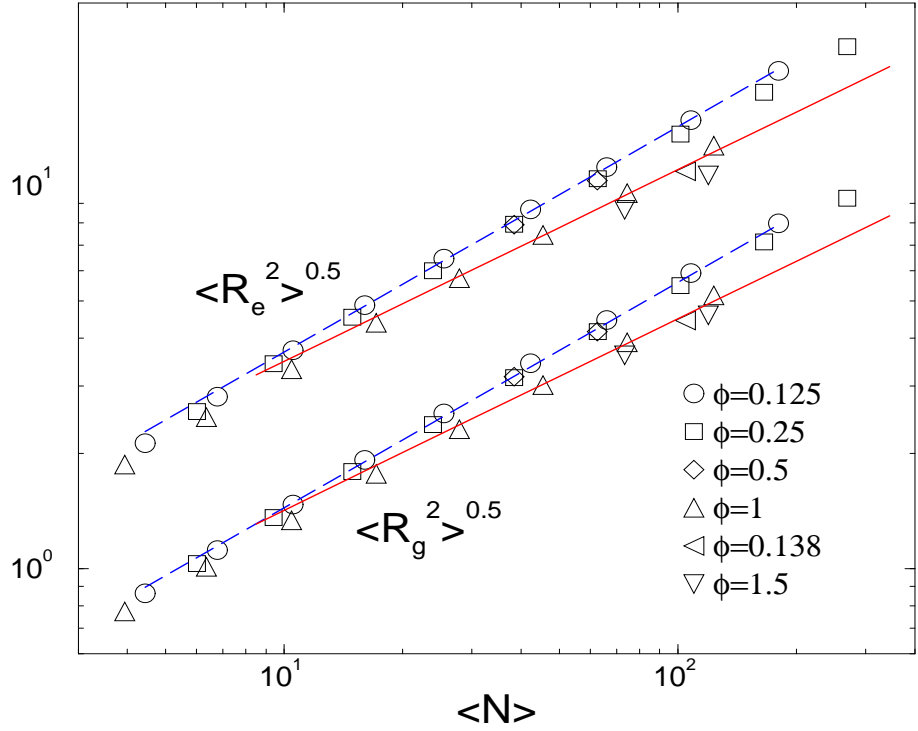,width=120mm,height=100mm}}
\caption{End-to-end distance \Rend\ and radius of gyration \Rgyr\
vs. mean chain mass \Nav. 
At small density ($\phi =0.125,0.25$) and, hence, weak chain overlap the chains 
are swollen (dashed line). At high density and mass we obtain Gaussian 
statistics (solid lines).}
\label{fig:RavNav}
\end{figure}

The averages over all chains of the mean-square end-to-end distance 
\Rend\ and the radius of gyration \Rgyr\ are plotted versus \Nav\ in Fig.\ref{fig:RavNav}.
Swollen and Gaussian scaling behavior are again obtained for low and strong chain overlap 
respectively. 
Note that, e.g., the chains for $\phi=1$ are Gaussian for $\Nav \geq 10$. 
Thus the position of a particular configuration $(E,\phi)$ with respect to
the crossover from dilute to semi-dilute regime can be determined. 
This is consistent with the value $\phi /\phistar$ given in Table \ref{tab:phiE}
where $\phistar=\Nav/(4\pi R_{g}^{3}/3)$ as usual.

\section{Molecular Weight Distribution}\label{sec:MWD}

We focus first on systems without rings, as in the rest of the paper, and
consider then the effects due to the ring closure constraint by allowing
linear chains and rings to compete.

\subsection{Systems without rings}\label{subsec:linear}

\begin{figure}[tbp]
\centerline{\epsfig{file=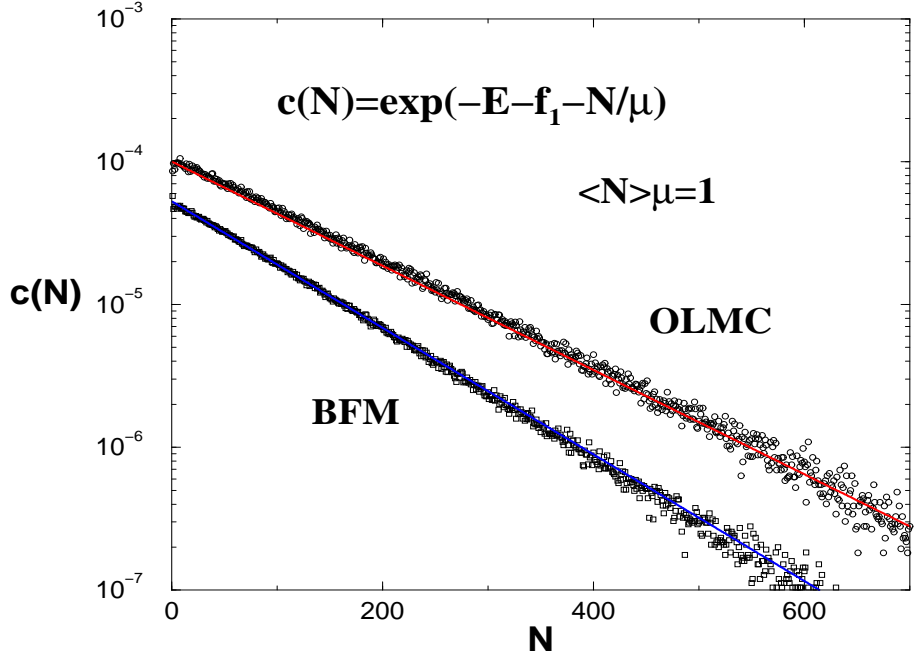,width=120mm,height=90mm}}
\caption{MWD $c(N)$ versus $N$ at high density for OLMC (upper curve; $E=7,\phi =1.5$) 
and BFM (lower curve; $E=7,8\phi =0.5$) systems.
Note that the distribution is always a pure exponential and {\em no} sign of
singularity was found for whatever density or energy. These curve are used
to fit \gL. For OLMC we have $\Nav\approx 119$, $\gL\approx 2.21$, 
for BFM $\Nav\approx 98$, $\gL\approx 2.85$.}
\label{fig:mwdhigh}
\end{figure}

Fig.~\ref{fig:mwdhigh} displays a typical MWD $c(N)$ obtained with the OLMC
algorithm at high density ($\phi =1.5$). This is compared with BFM data. The 
$c(N)$ is normalized such that $\phi =\sum_{N}Nc(N)$. Note that the free
monomers are counted as chains of length $N=1$. Both curves display to high
accuracy nice exponentials. This is a generic result for strongly
overlapping chains --- even at extremely high volume fractions. Notably, at
variance with a recent finding \cite{Yannick} no sign of singularity is
observed.

\begin{figure}[tbp]
\centerline{\epsfig{file=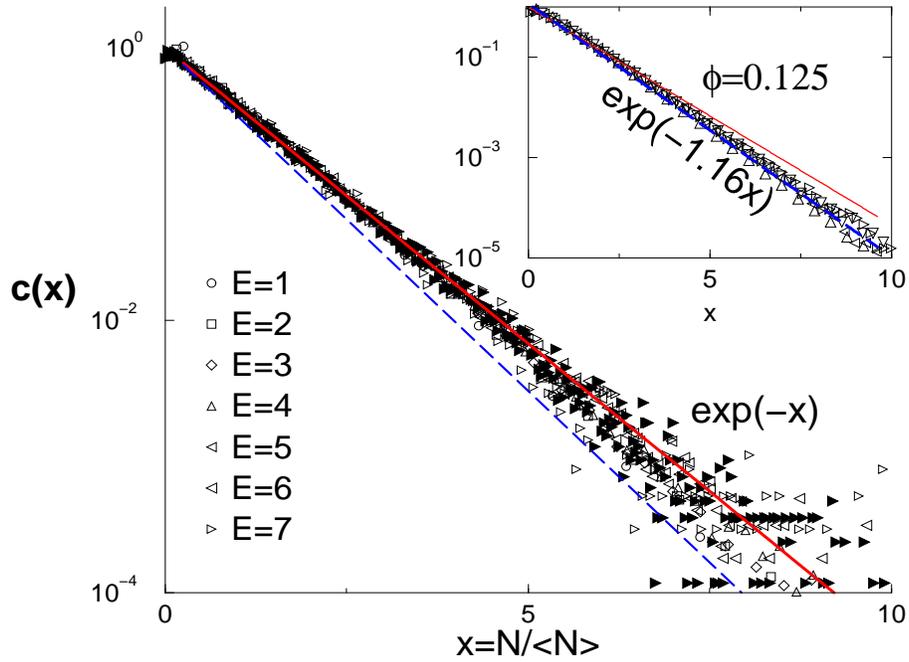,width=120mm,height=90mm}}
\caption{Data collapse of (properly normalized) MWD $c(x)$ versus $x=N/\Nav$ for OLMC. 
In the main figure the high density
scaling prediction $c(x)=\exp(-x)$ is verified (full line). The open symbols
denote density $\phi=1$ and the full symbols $\phi=1.5$.
Inset: The MWD in the dilute limit (OLMC at $\phi=0.125$) compares
well with the prediction $c(x) \propto \exp(-\gamma x)$ (dashed
line). The same symbols are used as in the main figure. }
\label{fig:mwdscal}
\end{figure}

In order to corroborate this result we try to scale the MWD obtained for
different densities and scission energies. In Fig.\ref{fig:mwdscal} we have
plotted the (properly normalized) MWD versus the natural scaling variable 
$x=N/\Nav$. In the high density limit (main figure)
data from two densities at various scission energies $E$ (as indicated in
the figure) collapses on a single 'master distribution' $c(x)=\exp (-x)$. In
the dilute limit, i.e. for the non-overlapping chains shown in the inset, we
find again a data collapse, but with slightly different slope in linear-log
coordinates. (Obviously, for large chain length the statistics deteriorates.)

How can these results be rationalized? Within a standard Flory-Huggins
mean-field approach \cite{com:GM} one may write the total free energy
density as 
\begin{equation}
\Omega \lbrack c(N)]=\sum_{N=1}^{\infty }c(N)\left( \log (c(N))+\mu N+%
\fchain(N,\phi,E)\right) .  \label{eq:MFhyp}
\end{equation}
The first term on the right is the usual translational entropy. The second
term entails a Lagrange multiplier which fixes the total monomer density $%
\phi $. The most crucial last term encodes the free energy of a reference
chain of length $N$ in the field created by the surrounding chains and free
monomers. \fchain\ will in general depend on the chain length $N$,
the density $\phi$ and the bonded and non-bonded interaction
parameters of the model studied \cite{com:fchaincN}. We have not computed
here irrelevant additive terms (such as virial terms) which are not
conjugated to $c(N)$. By functional derivation with respect to $c(N)$ one
readily obtains the equilibrium MWD: 
\begin{equation}
c(N)=\exp (-E-\fend(N,\phi ,E)-\mu N).  \label{eq:cN}
\end{equation}
We have defined here $\fend=\fchain-E-1$ and absorbed
all contributions to \fend\ linear in $N$ within the Lagrange
multiplier. Hence, within the mean-field approximation eq.(\ref{eq:MFhyp})
the MWD discussed above probes directly the free energy of an EP chain 
(besides the terms linear in $N$ mentioned above which are fixed by the imposed density)
which essentially ``renormalizes" the scission energy $E$.

We infer from Fig.\ref{fig:mwdhigh} and Fig.\ref{fig:mwdscal} that for our
EP chains \fend\ is to very high accuracy mass independent. Two
classical examples where this is indeed rigorously true are non-interacting
rigid rods and Gaussian chains \cite{com:GM}. (There \fend\ is
even independent of $\phi$.) If this is the case, the MWD becomes then a
pure exponential 
\begin{equation}
c(N)dN/\phi =c(x)dx=\exp (-x)dx  \label{eq:pureexp}
\end{equation}
where the scaling variable is $x=N/\Nav$, i.e. the
inverse Lagrange multiplier equals the mean-chain length, $\Nav \mu = 1$.


This is in agreement with our scaling in the high density limit which
implies that \fend\ is within numerical accuracy independent of $N$. 
We have explicitly checked that here the MWD scales like 
$c(N)\propto \exp(-E-\fend(\phi)-N/\Nav)$ where \fend\ only depends on $\phi$.
The slightly different slope found in the dilute regime suggests,
however, a weak mass effect. It is relatively simple to understand these
results from the standard theory of polymers in good solvent solutions \cite
{WMC98}. There the chemical potential of ends is given by 
$\fend =(\gamma -1)/\nuD \log (\xi )$ where $\xi $ is size of the 'blob', i.e. the
excluded volume correlation length for chains of length $N$ at density $\phi 
$.

In the semi-dilute limit we have 
$\xi \propto g^{\nuD}\propto \phi^{-\nuD/(3\nuD-1)}$\cite{Degennesbook} 
($g$ denotes the number of monomers within a
blob), hence 
\begin{equation}
\fend= (\gamma -1)/(3\nuD -1) \log (\phi/\phi_0) 
\label{eq:fendhighdensity}
\end{equation}
is independent of mass and scission energy. 
The constant reference density $\phi_0$ was introduced here for dimensional reasons.
Obviously, eq.(\ref{eq:fendhighdensity}) can strictly hold only in the
asymptotic limit of large blob sizes ($\phi \rightarrow 0$). Additionally, 
one can not exclude weak chain-length dependence of \fend\ at $N\approx 1$, 
but this becomes irrelevant in the scaling limit of $\Nav \gg 1$ \cite{com:largechain}.

In the opposite dilute limit the correlation length becomes density
independent and is given by the size of the chain $\xi =R\propto N^{\nu}$.
Hence, $\fend=(\gamma-1)\log (N)$ and the MWD becomes the
Schultz-Zimm form given in the Introduction (eq.\ref{eq:MWD_GM}). Note that
the $\gamma$-exponent in 3D is only slightly larger than its mean field value 
$\gamma =1$. Hence, the predicted power law depletion in the MWD for small $N$
is very weak and requires a relatively large mean chain mass $\Nav \gg 1$
This is why within the range of the parameters accessible essentially only the 
slightly different exponential tail $c(x)\propto \exp (-\gamma x)$ is visible.

Obviously, some of our simulations are in the intermediate regime ($\phi \approx \phistar$)
between these limiting cases (depicted by the two slopes in Fig.\ref{fig:mwdscal}). 
In order to characterize all the MWD with one parameter we fitted the tail of 
$\log (c(N))$ with $-E-\gL -\mu N$ where both $\mu$ and \gL\ are fit parameters. 
In the dense limit naturally one finds $\mu \Nav=\gamma_{eff}\approx 1$. 
As soon as the overlap decreases $\gamma _{eff}\rightarrow \gamma$. 
Note that the \gL-variation with $E$ for given $\phi$ is weak and we have only 
tabulated the value for the highest $E$, i.e. for the most strongly overlapping chains. 
The values are given in Tab.\ref{tab:phi} and are discussed in the subsequent 
Section\ref{sec:Nav}.

\subsection{Mixed systems with linear chains and rings}\label{subsec:rings}

\begin{figure}[tbp]
\centerline{\epsfig{file=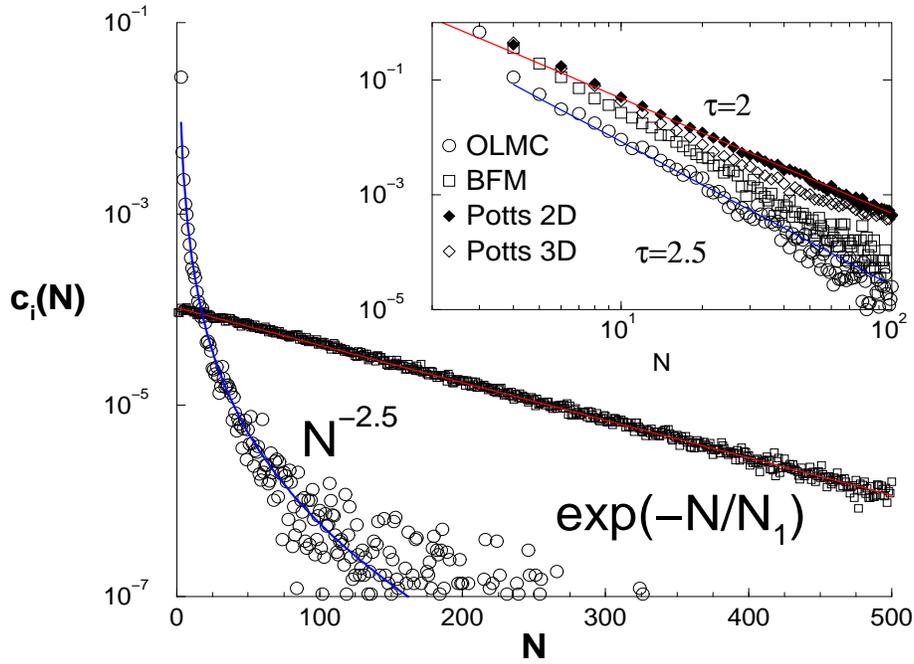,width=120mm,height=90mm}}
\caption{MWD for mixed systems containing rings. Main figure: \cR-MWD
of rings (spheres), \cL-MWD of linear chains (squares). Inset: Ring
MWDs for different models and dimensions. The indicated slopes confirm 
$\tau = 5/2$ in 3D and $\tau = 2$ in 2D. }
\label{fig:mwdring}
\end{figure}

First we briefly assess the importance of the ring closure constraint. We
allow the formation of rings so that both linear chains and rings may
coexist and compete for the monomers. In Fig.\ref{fig:mwdring} we present
the MWD of rings $\cR(N)$ and of linear chains $\cL(N)$ at density $\phi
=1.5$ and $E=7$. The linear chains appear to be unaffected by the presence
of rings and scale, as before, $\cL(N)\propto \exp (-\mu N)$ where the
slope $\mu $ is given by the inverse of the mean mass of the {\em linear}
chains $N_1$.
This can be easily understood by generalizing the Flory-Huggins expression
eq.(\ref{eq:MFhyp}) as a sum over the two different species $s=0,1$. The two
different MWD decouple and the functional derivation with respect to both
yields simply $c_{s}(N)=\exp (-f_{chain}^{s}(N,\phi ,E)-N\mu )$ for both
species in analogy with eq.(\ref{eq:cN}). The free energy of the linear
chain is $f_{chain}^{1}=E+1+f_{end}$ \ as before. In the strong overlap
regime $f_{end\text{ }}$ becomes again independent of mass $N$ ( hence, the
pure exponential seen for $\cL(N)$) and, as a matter of fact, equals the
density dependent value $f_{1}(\phi )$ obtained for the purely linear
systems which is tabulated in Tab.\ref{tab:phi}. Therefore the mean mass of
linear chains determines once again the chemical potential $\mu $.

However, our systems of {\em flexible} rings appear to be ring dominated,
i.e. much more monomers are contained in closed loops than in linear chains,
as expected. In contrast in systems of semi-flexible polymer chains the formation of
rings is going to be strongly suppressed \cite{SW99}. The MWD of rings is
not exponential, but rather shows a clearly pronounced power law behavior 
$\cR(N)\propto N^{-\tau }$ where $\tau =2.5$. (This does not exclude,
however, a final exponential cut-off, which can hardly be detected because
of the dominating power law.) The exponent $\tau $ can be obtained by a
simple argument due to Porte \cite{Porte}. The ratio
of both MWD gives the ratio of the two partition functions $Z_{s}$
\begin{equation}
\frac{\cR(N)}{\cL(N)}=\frac{Z_{0}(N)}{Z_{1}(N)}=\exp
(f_{chain}^{1}-f_{chain}^{0}).  \label{eq:ratiopart}
\end{equation}

The ratio $Z_{1}/Z_{0}$ is equal to the probability of opening a loop, which
must be proportional to: (i) the Boltzmann weight, $\exp (-E)$, due to the
constant scission energy, (ii) the number of places where the ring can
break, $N$, and (iii) the volume $R^{D}\propto N^{D\nu}$ that two
neighboring \ segments can explore after being disconnected ($D$ denotes the
dimension and $\nu$ the relevant Flory exponent at density $\phi $). Hence, 
$\cR\propto N^{-\tau}\exp(-N/N_1)$ with $\tau =D\nu +1=5/2$ 
at strong chain overlap in $D=3$, and $\tau =2$ in $D=2$ where $\nu=1/2$. 
Both exponents are well born out by the measured \cR-slopes 
depicted in the inset of Fig. \ref{fig:mwdring}. We compare here data
we obtained with three different algorithms: OLMC, BFM and Grand Canonical
Potts model (in 2D and 3D). The exponents in both dimensions are fully
consistent with the predicted ones.

We return now again to systems containing linear chains only.

\section{The scaling of the average chain mass}\label{sec:Nav}

So far we have reported on the general form and scaling of the MWD. Now we
want to go further and to investigate the scaling of the average chain mass 
and the second moment, i.e. the polydispersity index 
$I=\langle N^{2}\rangle /\langle N\rangle ^{2}$ , with regard to
density $\phi $ and scission energy $E$ (see Tab.\ref{tab:phiE}).


In the high density limit, i.e. within the validity of eq.(\ref{eq:pureexp}), 
the inverse Lagrange multiplier equals the mean-chain length as mentioned above
and the polydispersity $I=2$.
From the normalization constraint $\phi =\sum_{N}Nc(N)$ one infers that the 
mean mass is generally given by 
\begin{equation}
\Nav=\mu^{-1}=\sqrt{\phi \exp (E+\fend(\phi))}.
\label{eq:Navfend}
\end{equation}

For semi-dilute polymer chains one obtains from eq.(\ref{eq:Navfend}) and
eq.(\ref{eq:fendhighdensity}) $\Nav\propto \sqrt{\phi^{1+(\gamma -1)/(3\nuD-1)}\exp (E)}$ 
in agreement with the exponents $\alphaS\approx 0.6$ and $\deltaS=1/2$ quoted in
the Introduction. Similarly, one finds in the dilute limit 
$\alphaD = \deltaD = 1/(1+\gamma)\approx 0.46$ \cite{WMC98}.

\begin{figure}[tbp]
\centerline{\epsfig{file=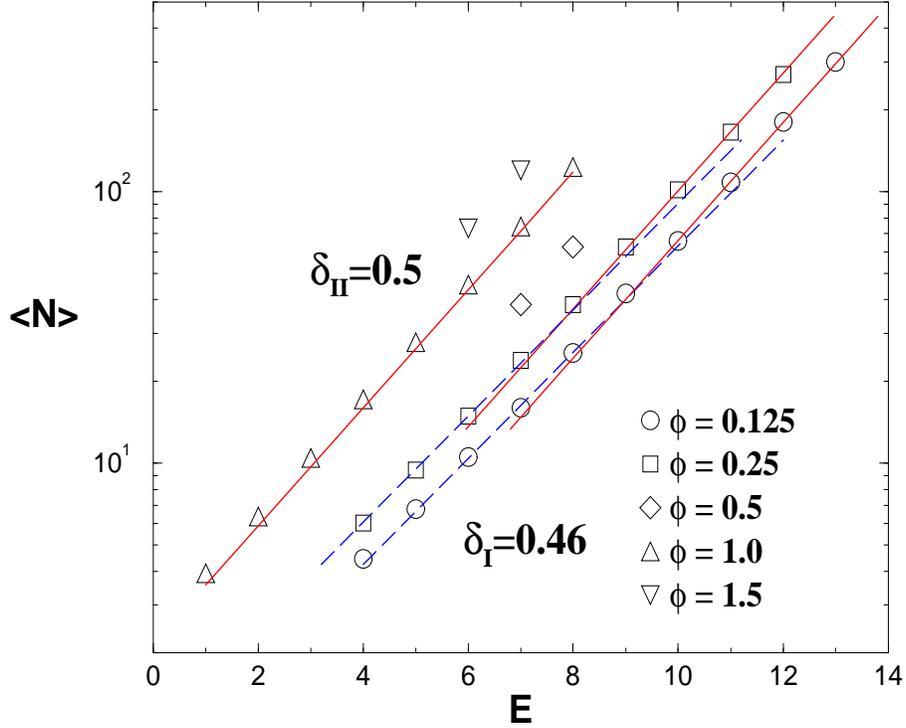,width=120mm,height=100mm}}
\caption{ Variation of mean chain mass \Nav\ with
dimensionless bond energy $E$ for various number densities $\phi$ as
indicated in the figure. At low chain overlap the data are consistent with
the exponent $\deltaD=0.46$ (dashed lines), at hight density
with the mean field exponent $\deltaS=0.5$ (full lines). }
\label{fig:NE}
\end{figure}

\begin{figure}[tbp]
\centerline{\epsfig{file=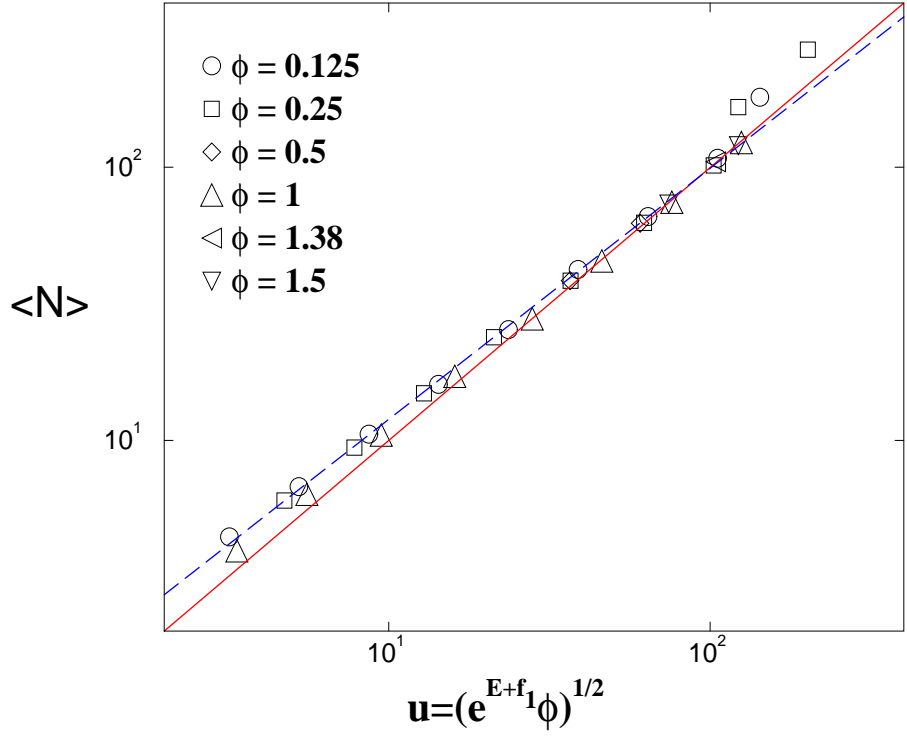,width=120mm,height=100mm}}
\caption{Scaling attempt in the high density limit for \Nav. 
The collapse confirms the scaling with respect to $u$ in the high chain 
overlap limit and the \gL\ obtained independently from the MWD $c(N)$.
Note that the remarkable good scaling works even in the dilute limit.
The dashed line marks for comparison the dilute exponent \alphaD.}
\label{fig:Nu}
\end{figure}

\begin{figure}[tbp]
\centerline{\epsfig{file=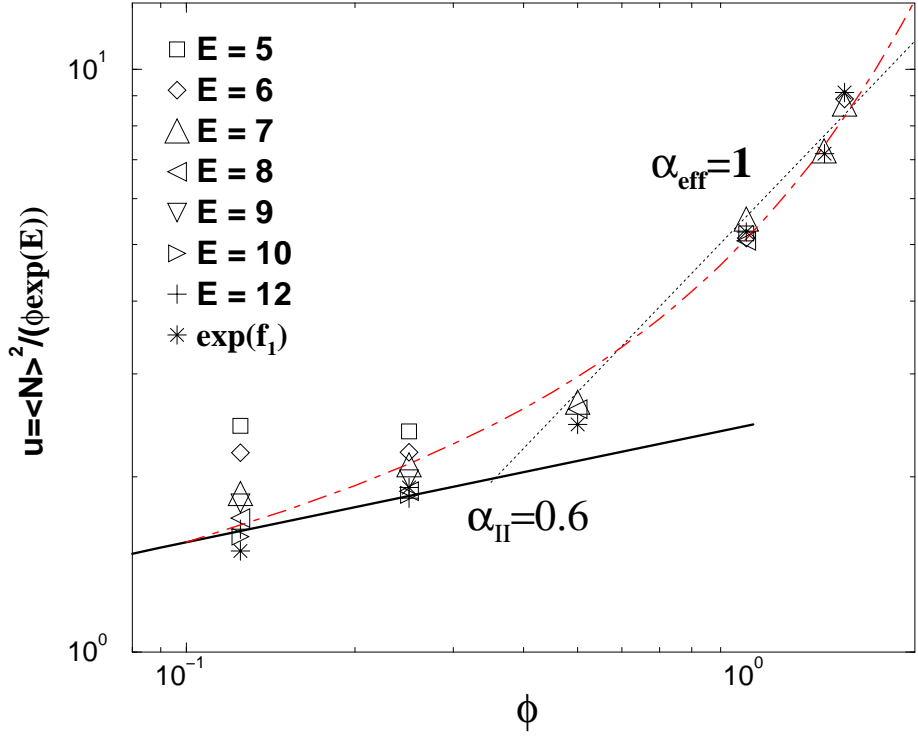,width=120mm,height=100mm}}
\caption{A crude rescaling attempt of mean chain mass $u=\Nav^2/(\phi\exp(E))$ 
versus $\phi$. Also indicated are the $\exp(\gL)$ estimated from the MWD (stars).
The mean chain mass rises faster for large densities $\phi \geq 0.5$. 
This regime is compared with an effective exponent $\alphaM \approx 1$ (dotted line)
and the nonalgebraic density dependence eq.(\protect\ref{eq:expansion})
with $B_{-1}=0.21$ and $B_1\approx 0.8$ (dashed-dotted line).}
\label{fig:Nphi}
\end{figure}

In Fig.\ref{fig:NE} we have plotted \Nav\ versus $E$
to check for the expected $\exp (\delta E)$ behavior. Despite the small
difference in both values one distinguishes for dilute systems a slope with $%
\deltaD=0.46$ and for the dense system ($\phi =1$) $%
\deltaS=0.5$. One can verify on Fig.\ref{fig:NE} that at
concentration $\phi =0.125$ a crossover into the semi-dilute regime for $%
E\approx 9$ occurs whereby the slope of the exponential function changes
from $\deltaD\approx 0.46$ to $\deltaS=1/2$, i.e.
the isolated polymer coils get large enough so they start touching each
other. At $\phi =0.25$ the concentration is already sufficiently high so
this happens at comparatively lower energies $E>5$. The $\deltaS=1/2$-slopes 
confirm that at high densities \fend\ becomes chain
length-independent and that eq.(\ref{eq:Navfend}) holds. This scaling
prediction is verified explicitly in Fig.\ref{fig:Nu} where we have plotted \Nav\
versus the scaling variable indicated by eq.(\ref{eq:Navfend}).
Here we have used the values $\gL(\phi,E)$ determined directly from the MWD. 
A plot using the prediction for semi-dilute polymers eq.(\ref{eq:fendhighdensity}) is, 
however, {\em not} successful for $\phi >0.5$. 
Hence, the measured \gL\ encapsulates physics other than the one expected in the limit of 
large blobs. To corroborate this further we plot in Fig.\ref{fig:Nphi} 
$u=\Nav^2/(\phi\exp(E))$ versus $\phi$. From eq.(\ref{eq:Navfend}) we know that
$u=\exp(\fend)$ for strong overlap. Not surprisingly the data collapse in that limit,
but not the dilute systems. (This will be improved in Fig.\ref{fig:Nuv}, see below.) 
Also indicated are the $\gL$ (stars) directly estimated from the MWD in the strong
chain overlap limit.
Notably, for $\phi \geq 0.5$ the mean mass appears to increase much faster than 
the growth exponent $\alphaS=0.6$. 
This breakdown of eq.(\ref{eq:fendhighdensity}) --- but not of eq.(\ref{eq:Navfend}) --- 
is not unexpected if one bears in mind that the scaling arguments can only be valid 
in the limit \cite{Degennesbook} of $\phi \rightarrow 0$ and $g\rightarrow \infty$, 
i.e. when the blob is larger enough.
This result suggests to rewrite eq.(\ref{eq:fendhighdensity})
as a systematic series expansion 
\begin{equation}
\fend(\phi) = B_{-1}\log(\phi/\phi_0) + B_{1} \phi + ...
\label{eq:expansion}
\end{equation}
where in order to match semi-dilute and melt regime we choose 
$B_{-1}=(\gamma -1)/(3\nuD-1) \approx 2.1$ and $\phi_0 \approx 0.018$.
The fit using eq.(\ref{eq:expansion}) with $B_1\approx 0.8$, depicted in Fig.\ref{fig:Nphi},
is only qualitatively satisfactory and further study is warranted. 
The above expansion is motivated by a recent second virial theory on
wormlike micelles and rigid rods interacting via  hard-core excluded volume \cite{SC94}.
The applicability of this theory to our systems of {\em flexible} EP 
with persistence length $b$ of order of the sphere diameter (see Tab.\ref{tab:phi}) 
appears to be unclear. Nevertheless, it is interesting to note that the second virial 
coefficient $B_1 = 8/3 \ v \approx 0.6$ predicted in \cite{SC94} is relatively 
close to the constant measured. Alternatively, the dependence of \Nav\ on $\phi$
might be represented by an effective growth exponent $\alphaM\approx 1$ (dotted line).
These values agree also favorably with an earlier lattice simulation \cite{Yannick}
although this is presumably accidental due to the different microscopic physics
which must intervene.

A proper crossover scaling between dilute and semi-dilute regimes (but not
with regard to the molten regime) is achieved by plotting 
$\Nav/\Nstar$ versus $\phi/\phistar$ in Fig.\ref{fig:Nuv}. 
The crossover length $\Nstar\propto \exp (-\varphi E)$ 
with $\varphi =(1/2-\alphaD)/(\alphaS-\alphaD)\approx 0.26$ 
and the crossover density  $\phistar\propto \exp (\kappa E)$ with 
$\kappa =\alphaD (1-\varphi )\approx 0.34$ are readily found by matching the asymptotic
behaviors of the dilute and semi-dilute regime. We have included data from
both OLMC and the BFM. For clarity, the BFM data have been shifted downwards
by a factor 10. As one anticipates from Fig.\ref{fig:Nphi}, only OLMC data
for $\phi \leq 0.5$, i.e. volume fractions smaller $0.1$ 
collapse properly on the predicted asymptotes ---
confirming hence the polymer physics expressed in eq.(\ref{eq:fendhighdensity}) 
--- while at higher densities simple liquids physics becomes relevant.

\section{Summary}\label{sec:Conclusion}

We have proposed here a new off-lattice Monte Carlo algorithm (OLMC) for
systems of equilibrium polymers (EP). It is shown that this model faithfully
reproduces the results of ref.\cite{WMC98}: the MWD of linear EP is
essentially exponential even in the limit of very high density in contrast
to recent claim\cite{Yannick}. Note that our findings do not support recent
claims of Levy-flight dynamic behavior in EP and GM which would require a
singularity in the MWD of linear chains \cite{Levyflight}.

If ring formation is allowed, however, the MWD for rings alone is strongly
singular, $\cR(N)\propto N^{-\tau }$ with $\tau =D\nu +1.$ This result has
been confirmed both in two- and in three dimensions for different lattice
and off-lattice models. We have also shown that the MWD of linear chains is
not affected by the presence of rings in the system which demonstrates that
the general Flory-Huggins approach is very accurate even in this case.

The mean chain length of linear EP is found to vary with density $\phi$ and
scission energy $E$ 
like $\Nav \propto \phi^{\alpha }\exp (\delta E)$ 
as observed in our earlier Bond Fluctuation Model investigation \cite{WMC98}
with exponents $\alphaD=\deltaD\approx 0.46$ in the dilute regime and 
$\alphaS=0.6,\deltaS=0.5$ in the semi-dilute regime. 
This holds only for volume fractions smaller $0.1$.

At higher densities when the blobs become too small the scaling approach
breaks down. Conformational properties are then largely determined by packing
effects as in simple liquids. The mean chain length then grows much faster
with $\phi$, qualitatively following a nonalgebraic dependence similar to 
a recent decription put forward for rod-like and semi-flexible micelles \cite{SC94}.
Our result is also compatible to an effective growth exponent $\alphaM\approx1$,
in agreement with recent simulations \cite{Yannick} and even with experimental 
observations \cite{Schurtenberger} although in the latter case this agreement might be
accidental in view of the rather low volume fraction probed experimentally.

As a further development the OLMC algorithm the model will be used for
studies of the dynamic \ behavior of linear and ring EP for which
investigations are currently underway.

\section*{Acknowledgments}

This research has been supported by the National Science Foundation, Grant
No. INT-9304562 and No. DMR-940518, and by the Bulgarian National Foundation
for Science and Research under Grant No. X-644/1996. JPW thanks J.L.~Barrat,
L.~Bellier-Castella, M.E. Cates, Y.~Rouault and P.~van~der~Schoot for 
stimulating discussions.

\begin{table}[t]
\begin{tabular}{|l|c||c|c|c|c|c|}
\hline
$\phi $ & $E$ & \Nav & $I$ & \Rend &  \Rgyr & $\phi/\phistar$ \\ \hline
0.125 & 8  & 25 & 1.86 & 6.4  & 2.5 & 0.3 \\
0.125 & 9  & 42 & 1.88 & 8.7  & 3.4 & 0.5 \\
0.125 & 10 & 66 & 1.89 & 12.2 & 4.5 & 0.7 \\ 
0.125 & 11 & 108 & 1.93 & 14.8 & 5.9 & 1.0 \\ 
0.125 & 12 & 181 & 1.95 & 19.9 & 8.0 & 1.5 \\
0.25 & 7 & 24 & 1.88   &  6.0  & 2.4 & 0.6\\
0.25 & 8 & 38 & 1.90   &  7.9  & 3.2 & 0.9 \\  
0.25 & 9 & 63 & 1.93 & 10.4 & 4.2 & 1.2 \\ 
0.25 & 10 & 102 & 1.93 & 13.6 & 5.5 & 1.7 \\ 
0.25 & 11 & 166 & 1.92 & 17.5 & 7.1 & 2.3 \\ 
0.25 & 12 & 270 & 1.99 & 23.1 & 9.3 & 3.1 \\ 
0.5  & 7  & 38.4& 1.92 & 7.9  & 3.2 & 1.8 \\
0.5 & 8 & 62 & 1.94 & 10.3 & 4.2 & 2.5 \\
1 & 5 & 28 & 1.92 & 5.8  & 2.3 & 1.9 \\
1 & 6 & 46 & 1.94 & 7.5  & 3.0 & 2.5 \\
1 & 7 & 74 & 1.94 & 9.6 & 3.9 & 3.4 \\ 
1 & 8 & 123 & 1.96 & 12.7 & 5.2 & 4.8 \\ 
1.38 & 7 & 105 & 1.97 & 10.9 & 4.5 & 5.0 \\ 
1.5 & 6 & 73 & 1.95 & 8.6 & 3.6 & 4.0 \\ 
1.5 & 7 & 120 & 1.97 & 10.6 & 4.6 & 5.1 \\ \hline
\end{tabular}
\vspace*{0.5cm}
\caption{Summary of measured quantities for configurations with $\Nav > 20$.
Quantities tabulated: the mean chain mass \Nav, the
polydispersity index $I=\langle N^{2}\rangle /\Nav^{2}$, 
the mean end-to-end distance \Rend\, 
the mean gyration radius \Rgyr, and the chain overlap $\phi/\phistar$.
Note that $I$ increases systematically with chain overlap.}
\label{tab:phiE}
\end{table}

\begin{table}[t]
\begin{tabular}{|l||c|c|c|c|}
\hline
$\phi $ & $l$   	& $b$  	&  $\xi$& \gL \\ \hline
0.125 	& 0.7574  	& 1.41 	& 4.7	& 0.4 \\ 
0.25 	& 0.7565   	& 1.33 	& 4.0	& 0.65 \\ 
0.5 	& 0.7530    	& 1.2 	& 3.8	& 0.9 \\ 
1.0 	& 0.7408    	& 1.12 	& 3.3	& 1.66 \\ 
1.38 	& 0.7234   	& 1.05 	& 2.8	& 1.97 \\ 
1.5 	& 0.7158    	& 1.02 	& 2.5	& 2.21 \\ \hline
\end{tabular}
\vspace*{0.5cm}
\caption{Density dependence of mean bond length $l=\bb^{1/2}$, 
the persistence length $b$ for Gaussian chains of blobs,
the blob size $\xi$ estimated from the slopes in Fig.\protect\ref{fig:RN}, 
and the free energy factor $\gL(\phi)$ 
obtained from the exponential tail of the MWD as explained in Sec.\ref{sec:MWD}.}
\label{tab:phi}
\end{table}

\newpage 

\newpage 

\newpage 

\newpage 

\newpage 

\newpage 

\newpage 

\newpage 

\newpage 

\newpage 
\begin{figure}[tbp]
\centerline{\epsfig{file=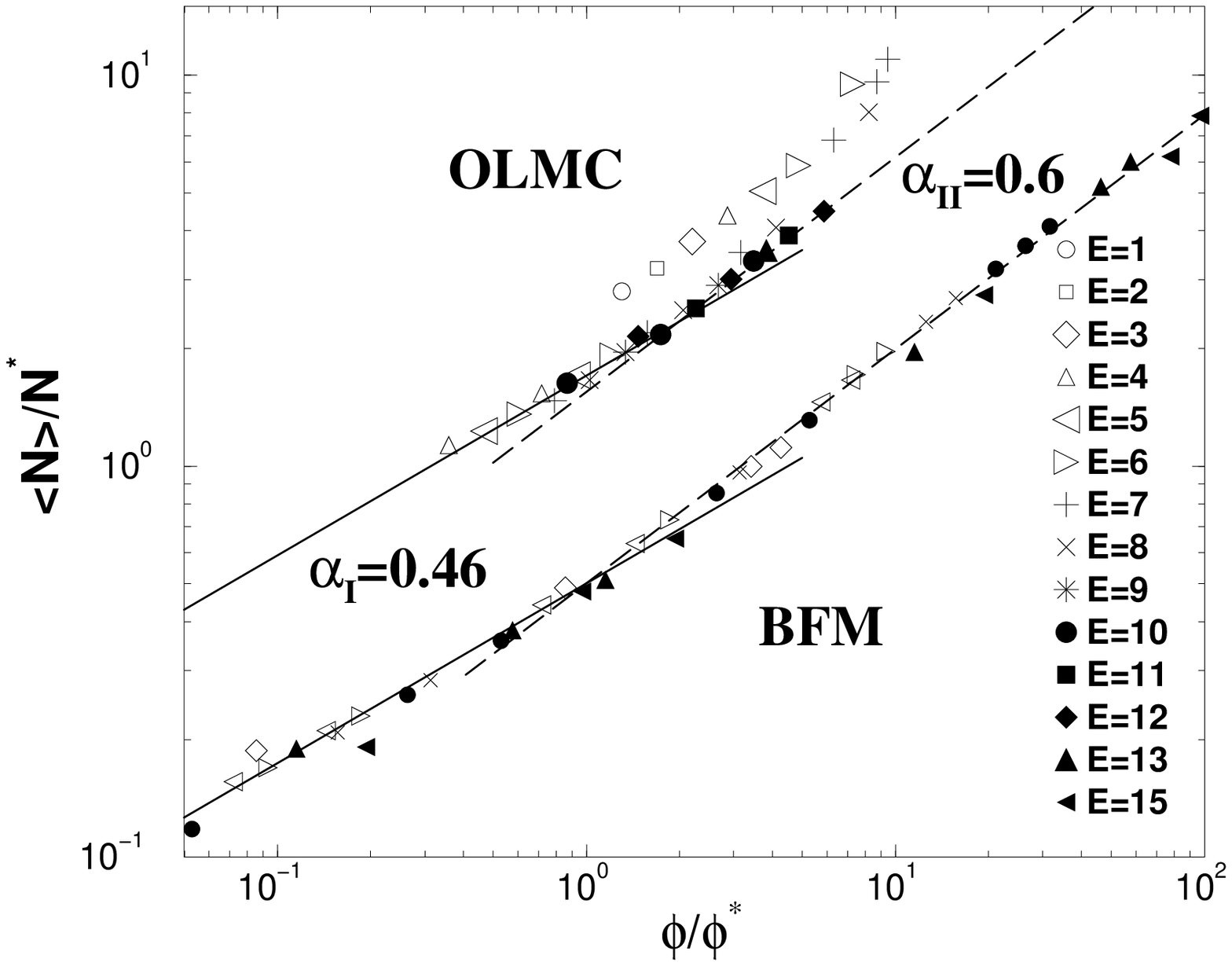,width=120mm,height=100mm}}
\caption{ Dilute-Semidilute crossover scaling $\Nav/\Nstar$ vs. $\phi/\phistar$ 
for OLMC and BFM (shifted downwards for clarity). For densities $\phi \geq 0.5$ the
OLMC data do not scale and are systematically above the semi-dilute
asymptote. This is due to additional entropic interactions between the dense
beads which increases the mean chain mass. }
\label{fig:Nuv}
\end{figure}


\begin{references}
\bibitem{com:fchaincN}  More generally, \fchain\ might even
depend explicitly on $c(N)$. This is of relevance in situations such as LP
brushes where the MWD is conjugated with the density profile of the layer 
\cite{MWL99b}.
\bibitem{com:largechain}  Minor corrections to Eq.(\ref{eq:fendhighdensity})
from the small fraction of chains that are too long for their
excluded-volume interactions to be screened by the surrounding chains (under
melt conditions this applies \cite{Degennesbook} to those chains whose
length exceeds $\langle N\rangle ^{2}$). However, this contribution is
exponentially small and can be neglected.
\end{references}
\end{document}